\documentclass[prd,preprint,superscriptaddress]{revtex4}
\usepackage{revsymb}
\usepackage{amsmath}
\usepackage{graphicx}
\newcommand{\be}{\begin{equation}}
\newcommand{\ee}{\end{equation}}
\newcommand{\ben}{\begin{eqnarray}}
\newcommand{\een}{\end{eqnarray}}
\newcommand{\vphi}{\varphi}
\newcommand{\n}{\label}
\newcommand{\no}{\noindent}

\newcommand{\ga}{\gamma}

\begin{document}

\title{Dual interacting cosmologies and late accelerated expansion}
\author{Luis P. Chimento\footnote{Electronic Mail Address: chimento@df.uba.ar}}
\affiliation{Departamento de F\'{i}sica, Facultad de Ciencias
Exactas y Naturales, Universidad de Buenos Aires, Ciudad
Universitaria, Pabell\'{o}n I, 1428 Buenos Aires, Argentina}
\author{Diego Pav\'{o}n\footnote{Electronic Mail Adress: diego.pavon@uab.es}}
\affiliation{Departamento de F\'{\i}sica, Facultad de Ciencias,
Universidad Aut\'{o}noma de Barcelona, 08193 (Bellaterra)
Barcelona, Spain}
\begin{abstract}
In this paper we show that by considering a universe  dominated by
two interacting components a superaccelerated expansion can be
obtained from a decelerated one by applying a dual transformation
that leaves  the Einstein's field equations invariant.
\end{abstract}
\maketitle

\section{Introduction}
Currently, the view  that the Universe has entered a stage of
accelerated expansion is widely shared by cosmologists \cite{Adam}
to the point that the debate has shifted to discussing when the
acceleration did actually commence and if it is just a transient
phenomenon or it will last forever and, above all, which is the
agent behind it. Whatever the latter, usually called dark energy,
it must possess a negative pressure high enough to violate the
strong energy condition (SEC). A number of dark energy candidates
obeying the dominant energy condition (DEC) have been proposed,
ranging from an incredibly tiny cosmological constant to a variety
of exotic fields (scalar, tachyon, {\it k}-essence, and so on)
with suitably selected potentials -see Ref. \cite{reviews} for
reviews. However, observations seem to marginally favor some or
other energy field -dubbed ``phantom energy"- that violates the
DEC \cite{caldwell} over dark energy fields that satisfy it.
Likewise, lately, it has been shown the existence of dual symmetry
transformations that leaves invariant the Einstein field equations
for spatially flat, homogeneous and isotropic universes
\cite{lpch1}. These transformations prove themselves extremely
useful since they allow to obtain phantom dominated expansions
from contracting scenarios \cite{lr1,jlr,jel,lpch2}. Other
features of phantom cosmologies have been investigated in
\cite{todos}.

The aim of this paper is twofold: $(i)$ To extend the technique of
dual symmetry transformations that preserve the form of Einstein's
equations to the case that the expansion of the Universe is
dominated by two fluids (dark matter and dark energy) that
interact with each other. The dark energy fluid may be of phantom
type or not. $(ii)$ To apply this technique to three cases in
which the dark energy is a different phantom fluid.

In section II we sketch the dual symmetry transformation when both
fluids are noninteracting and then extend the transformation first
to the case that they interact and both of them satisfy the DEC,
and then to the case that one of them does not satisfy the DEC.
Likewise, we study the evolution of the ratio between both energy
densities. It turns out that when the interaction term is
proportional to the total energy density the aforesaid ratio tends
asymptotically to a constant. In section III we apply the method
of section II, successively, to the cases that the phantom
component is a scalar field with negative kinetic energy, a {\it
k}-essence field and a tachyon field. Finally, in section IV we
summarize our conclusions and present some comments.

\section{Dual symmetry for interacting fluids scenarios}
Let us consider a homogeneous, isotropic and spatially flat
universe filled by two fluids of energy densities and pressures
$\rho_{i}$ and $p_{i}$ (with $i = 1, 2$), respectively. The
Friedmann equation and the energy conservation equation read
\ben
3 H^{2}& = & \rho_{1} + \rho_{2}\, ,
\nonumber \\
\dot{\rho}_{1} + &\dot{\rho}_{2}& + \; \, 3 H (\rho_{1}+\rho_{2}+
p_{1} + p_{2}) = 0 \, ,
\label{friedmann1} \een
\\
where we have set $c = 8\pi \, G =1$.

It can be readily seen that in this rather general scenario there
is a dual symmetry relating this cosmology to another one (with
two fluids of energy densities and pressures $\bar{\rho}_{i}$, and
$\bar{p}_{i}$), generated by
\\
\ben \bar{\rho}_{1} &=&\alpha\rho_{1} + (1 -\beta) \rho_{2}\, ,
\nonumber \\
\bar{\rho}_{2} &=& (1 - \alpha)\rho_{1} + \beta \rho_{2}\, ,
\nonumber \\
\bar{H}& = & - H \, ,
\label{dual1}
\een
\\
where the parameters of the transformation,
\[
\alpha =
\frac{\bar{\gamma}_{2}+\gamma_{1}}{\bar{\gamma}_{2}-\bar{\gamma}_{1}}\,
\qquad \mbox{and} \qquad \beta =
-\frac{\gamma_{2}+\bar{\gamma}_{1}}{\bar{\gamma}_{2}-\bar{\gamma}_{1}}\,
,
\]
solely depend on the barotropic indexes of the fluids. As usual,
these indexes are given by $\gamma_{i} =1+(p_{i}/\rho_{i})$ and
parallel expressions for the $\bar{\gamma}_{i}$ of the other
cosmology. We define the overall barotropic index
$\gamma = (\gamma_{1}\, \rho_{1}\, + \,\gamma_{2}\,%
\rho_{2})/(\rho_{1}\,+ \, \rho_{2})$ for the unbarred cosmology.
An entirely parallel expression exists for $\bar{\gamma}$ in the
other cosmology. Obviously the duality transformation connects
these two indexes by $\bar{\gamma} \rightarrow -\gamma$. This
means that $\rho_{1}+\rho_{2}+p_{1}+p_{2} \rightarrow
-(\rho_{1}+\rho_{2}+p_{1}+p_{2})$. Put another way, if the DEC is
fulfilled in one cosmology, then it is violated in the other.
The transformation law (\ref{dual1}.c) implies $\bar{a}%
=1/a$. Accordingly, if one cosmology (say, the unbarred one)
describes a phase of contraction, the barred one describes a phase
of expansion, i.e.,  both cosmologies are dual of each other
\cite{lpch1}.

In the remainder of this section we generalize this technique to
the case that the fluids do not conserve separately but interact
with each other and then investigate the consequences. We begin by
writing
\\
\ben
3 H^{2}& = & \rho_{1} + \rho_{2}\, ,
\nonumber \\
\dot{\rho}_{1} + 3&H& \gamma_{1}\, \rho_{1}= -3H\, \Pi\, ,
\nonumber \\
\dot{\rho}_{2} + 3&H& \gamma_{2}\, \rho_{2}= 3H \, \Pi\, ,
\label{friedmann2}
\een
\\
where the quantity $\Pi$ characterizes the interaction.
Automatically  the above dual symmetry gets restricted to the
following transformation $\rho_{i}\rightarrow \rho_{i}$, $H
\rightarrow -H$, $\gamma_{i} \rightarrow -\gamma_{i}$, and $\Pi
\rightarrow - \Pi$, with the
overall barotropic index transforming as $\gamma%
\rightarrow-\gamma$. Therefore, there is a duality between two
cosmologies, driven by two interacting fluids through  the set of
equations (\ref{friedmann2}), that have the sign of the individual
barotropic indexes reversed. This opens the possibility of
considering phantom dark energy with a negative barotropic index,
which characterizes a ghost or phantom cosmology, as a source of
Einstein's equations.

Defining the energy density ratio $r = \rho_{1}/\rho_{2}$ and
using Eqs. (\ref{friedmann2}), we obtain the evolution equation
\\
\ben \label{.r} \dot{r} = -3\Gamma H\,r, \qquad
\Gamma=\gamma_{1}-\gamma_2+\frac{\rho_{1} \, + \,
\rho_{2}}{\rho_{1}\rho_{2}}\,\, \Pi. \een
\\
Since, except for the sign, dual cosmologies share the same
interaction term $\Pi$ the transformation $\Gamma \to%
-\Gamma $, holds. In addition, equation (\ref{.r}) and the ratio
$r = \rho_{1}/\rho_{2}$ are invariant under the dual
transformation thereby $r$ is a well defined quantity. In
particular if $\dot{r}$ vanishes in one cosmology, it also
vanishes in the dual one, meaning that the stationary solutions
$r=r_{s}$ of Eq. (\ref{.r}) are shared by both cosmologies.

Let us now assume that both fluids satisfy the DEC, $\rho_{i} + %
p_{i} > 0$ (i.e., none of them is of phantom type) but one of them
(say, fluid $2$) violates the SEC, $\rho_{2} + 3p_{2} < 0$ (i.e.,
it is a dark energy fluid), while the other does not, and
specialize the interaction term to $\Pi%
=-c^{2}(\rho_{1}+\rho_{2})$ with $c^{2}$ a small dimensionless
constant. This particular choice of $\Pi$ has proved interesting
because it provides analytical solutions and leads to a fixed
ratio matter/dark-energy at late times whatever the initial
conditions (see, e.g., \cite{ladw}, \cite{srd}). Farther ahead in
this Section we shall see that this is also true when the dark
energy is of phantom type.

The stationary solutions of Eq. (\ref{.r}) are obtained by solving
$r_s\Gamma(r_s)=0$. When $\gamma_{1}$ and $\gamma_{2}$ are
constants these solutions are given by the roots of the quadratic
equation
\\
\be \n{r0pm} r_s^\pm=-1+2b\pm 2\sqrt{b(b-1)}, \qquad
b=\frac{\gamma_{1}-\gamma_{2}}{4c^2} > 1 \, .
\ee
\\
These satisfy the inequalities $r_{s}^{+}\geq 1 \geq r_{s}^{-}$
and for this model the general solution of Eq. (\ref{.r}) read
\\
\be r(x)=\frac{r_s^- + xr_s^+}{1+x} \, ,
\label{rg}
\ee
\\
where $x=(a/a_0)^{-\lambda}$ with $\lambda\equiv%
12c^2\sqrt{b(b-1)}$. It is readily seen that $r(x)$ is a monotonic
decreasing function in the range $r_{s}^{-} < r <r_{s}^{+}$.
Finally, near this attractor solution, $r \approx r^{-}_{s}$, the
last two equations (\ref{friedmann2}) can be approximated by
\\
\ben \n{1'}
\frac{\rho_{1}'}{\rho_{1}} \simeq \frac{\gamma_{1}-
c^{2}(1+1/r^{-}_{s})}{c^{2}\, (r^{+}_{s}-r^{-}_{s})(r-r^{-}_{s})},\\
\frac{\rho_{2}'}{\rho_{2}} \simeq \frac{\gamma_{2}+
c^{2}(1+r^{-}_{s})}{c^{2}\, (r^{+}_{s}-r^{-}_{s})(r-r^{-}_{s})},
\n{2'}
\een
\\
where the prime denotes derivative with respect to $r$. For nearly
constant barotropic indexes, $\gamma_{1}$ and $\gamma_{2}$, last
equations integrate to
\\
\be
\rho_{1} \propto%
a^{-3\left[\gamma_{1}-c^{2}(1+1/r^{-}_{s})\right]}, \quad
\rho_{2} \propto%
a^{-3\left[\gamma_{2}+c^{2}(1+r^{-}_{s})\right]}\, ,
\label{r12}
\ee
\\
while from the Friedmann equation (\ref{friedmann2}.a) the time
dependence of the scale factor
\\
\be a \propto (\pm\,\, t)^{\frac{2}{3[\gamma_{2}+
c^{2}(1+r^{-}_{s})]}}\,
\label{scalef}
\ee
\\
is readily obtained.

From the condition $\Gamma(r_s^-)=0$ it follows that the exponents
in the energy densities (\ref{r12}), which can be considered as
effective barotropic indexes, coincide. This shows that the
interaction modifies the apparent physical properties of the
fluids.

\begin{figure}[th]
\includegraphics[height=4.0in,width=5.0in,angle=-90,clip=true]{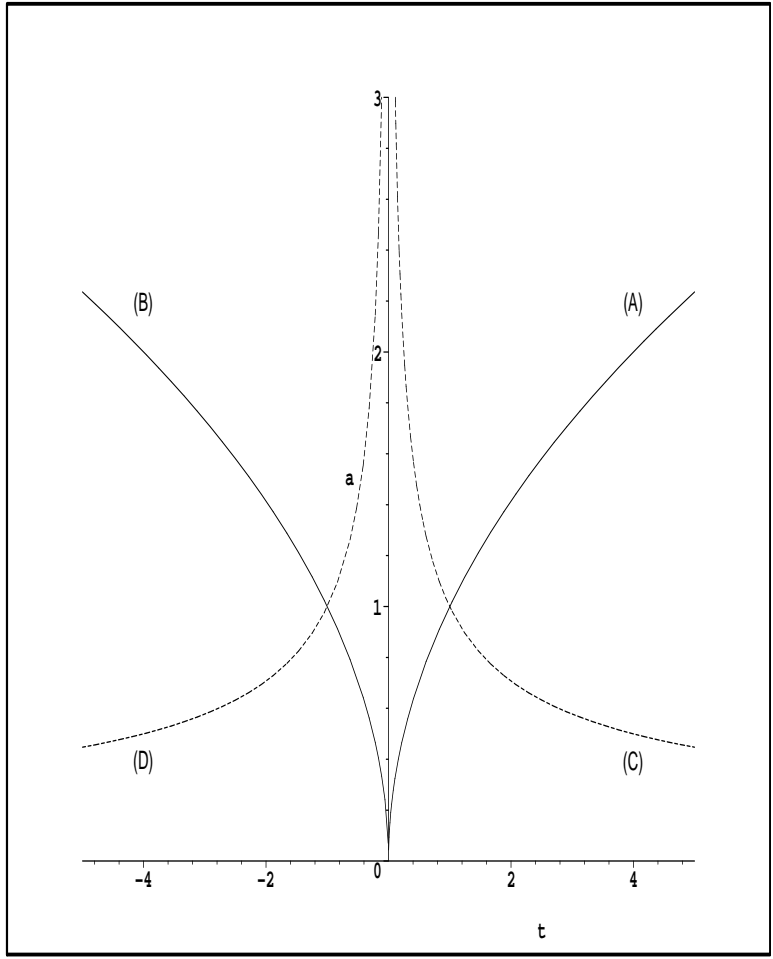}
\vspace{0.0cm} \caption{The four branches  of Eq. (\ref{scalef}).
The duality transformation  maps curve A into C (and viceversa).
Likewise, it maps curve B into D (and viceversa). Thus, curves A
and C are dual of each other, the same is true for the pair B, D
-see the text. The vertical axis corresponds to the scale factor
$a$.} \label{branches}
\end{figure}

We now apply this model to the case that the fluid $2$ violates
the DEC -i.e., it is a phantom fluid with $\gamma_{2}<0$. From the
two last expressions (\ref{r12}), (\ref{scalef}) and duality four
distinct possibilities emerge (see Fig. \ref{branches}):

\noindent $(i)\,$ $\gamma_{2}+c^{2}(1+r^{-}_{s})>0$,

\no for $t\ge 0$, the Universe expands from an initial singularity
at $t=0$ with a vanishing scale factor, $(A)$,

\no for $t\le 0$, the Universe contracts from the far past and
ends in a big crunch at $t=0$, $(B)$.

\noindent $(ii)\,$ $\gamma_{2}+c^{2}(1+r^{-}_{s})<0$ (the dual of
$(i)$, namely, $\gamma_{2}\rightarrow -\gamma_{2}$ and $c^{2}
\rightarrow -c^{2}$),

\no for $t\ge 0$ the Universe contracts from an initial
singularity at $t=0$ with an infinite scale factor, $(C)$,

\no for $t\le 0$, the Universe expands from the past and ends in a
big rip at $t=0$, $(D)$.

We have assumed, without loss of generality,  that $r$ is near the
attractor $r_{s}^{-}$; this facilitates the qualitative
description and more readily illustrates the dual symmetry.

We wish to emphasize that one may get a superaccelerated expanding
phase (i.e., $H>0$ together with $\dot H>0$) when
$|\gamma_2|>c^2(1+r_{s}^{-})$ and also when
$|\gamma_2|<c^2(1+r_s^-)$. In the latter case the superaccelerated
expanding phase is obtained by a dual transformation that reverses
the signs of $\gamma_{1}$, $\gamma_{2}$ and $c^{2}$. This
interchanges the roles of both fluids and replaces the term $\Pi$
by $-\Pi$.

\section{Phantom dark energy}
In this section we apply the above method to three specific cases
in which one component is matter (i.e., it satisfies the SEC) and
the other component is a phantom fluid (as such, it does violate
the SEC and DEC). For the latter we will consider in turn a scalar
field, a {\it k}-essence field and a tachyon field.

\subsection{Scalar field cosmology}
Let be an accelerated universe whose source of dark energy is a
scalar field $\vphi$ of phantom type. This type of fields may
arise in string theory, see \cite{frampton} and references
therein. We write its pressure and energy density admitting both
signs for kinetic energy term, see e.g. Refs. \cite{sami} and
\cite{lr2},
\\
\be
\rho_{\vphi}= s \textstyle{1\over{2}} \dot{\vphi}^{2} +%
V(\vphi), \qquad   p_{\vphi} = s\textstyle{1\over{2}}%
\dot{\vphi}^{2} -V(\vphi),
\label{rho&p}
\ee
\\
where $s$ is a constant that may bear either sign. It follows that
\\
\be
\n{s}
\gamma_{\vphi} = s%
\frac{\dot{\vphi}^{2}}{\rho_{\vphi}}\, .
\ee
\\
From the above equation we see that the barotropic index becomes
negative in two separate cases, viz, when $s<0$ with a real scalar
field and when $s>0$ with an imaginary scalar field. The dynamic
equations of both interacting components are
\ben
\dot{\rho}_{m} + 3H\gamma_{m}\, \rho_{m} = -3H\,\Pi\, \, , \nonumber\\
\dot{\rho}_{\vphi} + 3H\gamma_{\vphi}\, \rho_{\vphi} = 3H \, \Pi
\, , \label{consv} \een where $\rho_{m}$ indicates the matter
energy density. Since  $\rho_{m}$ and $\rho_{\vphi}$ may be seen
as functions of $r=\rho_{m}/\rho_{\vphi}$, with the help of
(\ref{.r}), Eq. (\ref{consv}.b) can be written as $\Pi=-r \Gamma
\rho'_{\vphi} +\gamma_{\vphi} \rho_{\vphi}$ and in accordance with
Eqs. (\ref{.r}) and (\ref{rho&p}) we obtain a differential
equation for the potential
\\
\be \frac{\Pi}{\rho_{\vphi}}= \gamma_{\vphi}- \Gamma \, r \,
\left[\frac{\gamma'_{\vphi}}{2-\gamma_{\vphi}}+
\frac{V'(\vphi)}{V(\vphi)}\right] \, .
\label{Pirho}
\ee
\\
The latter is very useful because when all the quantities that
enter it, except $V(\vphi)$ and $ V'(\vphi)$, are known functions
of the ratio $r$ the potential $V(r)$ can be obtained by
integration. Combining it with $r(a)$, derived from (\ref{.r}), we
can resort to the Friedmann's equation, $3H^2=\rho_{\vphi}(1+r)$,
to obtain the scale factor as a function of time. Also, in virtue
of the relation $\Pi = -c^{2} (\rho_{m}+ \rho_{\vphi})$, the
conservation equation (\ref{consv}.b) for $\varphi$ can be written
as
\\
\be \n{kg}
\ddot\vphi+3H\dot\vphi\left[1+\frac{c^2(1+r)}%
{\gamma_\vphi}\right]+\frac{1}{s}\frac{dV}{d\vphi}=0.
\ee
\\
Near the attractor dominated regime,  $r\approx r_{s}^{-}$, and
for constant barotropic indexes, $\gamma_m$ and $\gamma_\vphi$,
the scale factor has the power law solution $a \propto%
t^{2/3\nu_\vphi}$, given by (\ref{scalef}), with
$\nu_\vphi=\gamma_{\vphi}+c^2(1+r_{s}^-)$. In this approximation,
the simultaneous solution of Friedmann's equation  and (\ref{kg})
leads to a potential that can be cast as a series expansion in the
exponential potential (see Ref. \cite{ladw}). Approximating
$V(\vphi)$ by this term we have
\\
\ben
\n{po}
V(\vphi)\approx\frac{2(2-\gamma_\vphi)}{3\nu_\vphi^2(1+r_s^-)}\,
e^{-sA\vphi},
\\
\n{ca}
\vphi\approx \frac{2}{sA}\ \, \ln t \, , \qquad
A=|\nu_\vphi|\sqrt{\frac{3(1+r_s^-)}{s\gamma_\vphi}} \, .
\een
\\
When $s$ is negative, the parameter $A$ is real, so both the
phantom scalar field and potential become real quantities. By
contrast, when $s$ is positive, the parameter $A$ is imaginary
whence the phantom scalar field becomes imaginary but the dominant
term of the potential remains real. In general, applying the dual
transformation \be \n{tv} \bar V=s\dot\phi^2+V, \qquad \qquad
{\dot{\bar\phi}}^2=-\frac{s}{\bar s}\dot\phi^2, \ee to the
solution (\ref{po}), (\ref{ca}) we get the transformed potential
and scalar field for any $s, \bar s$ values. This transformation
together with the change of the interaction term $\Pi\to-\Pi$
reverses the sign of $\nu_\vphi$ and the new configuration is
given by the barred quantities.

\subsection {{\it K}-essence cosmology}
Here we consider the case in which the dark energy is provided by
a {\it k}-essence field, $\phi$, characterized by the Lagrangian
$\it{L}=-U(\phi)F(x)$. The potential $U(\phi)$ is a positive
definite function of the {\it k}-essence field $\phi$ and $F$
depends on the variable $x \equiv {g^{ik}\phi_i\phi_k}$ with
$\phi_{i} \equiv \partial\phi/\partial x_{i}$. This field arises,
for instance, in open bosonic string field theory \cite{fradkin}.
Identifying the energy-momentum tensor of the k field with that of
a perfect fluid, its energy density and pressure are given by \be
\n{B} \rho_\phi = U(F - 2xF_x), \qquad p_{\phi}= -U F \, , \ee
where the subscript $x$ means $d/dx$.

Assuming that this perfect fluid  obeys the barotropic equation of
state it follows that $ \gamma_\phi= -2xF_{x}/(F-2xF_{x})$, and $
\rho_\phi= UF/(1-\gamma_{\phi})$. The {\it k}-essence field
represents phantom dark energy when $\gamma_\phi$ is negative.
This requires a decreasing, positive-definite kinetic function.
The Friedmann and the conservation equation for the {\it
k}-essence field can be written as \be \n{00k}
3H^2=\frac{UF(1+r)}{1-\gamma_\phi}, \qquad
r=\frac{\rho_m}{\rho_\phi}, \ee \be \n{kgk}
[F_x+2xF_{xx}]\ddot\phi+3HF_x\dot\phi\left[1+%
\frac{c^2(1+r)}{\gamma_\phi}\right]+\frac{V'}{2V}[F-2xF_x]=0. \ee

Again, near the attractor dominated regime and for constant
$\gamma_m$ and $\gamma_\phi$, the scale factor has the power law
solution $a \propto t^{2/3\nu_\phi}$ given by Eq. (\ref{scalef}),
with $\nu_\phi=\gamma_{\phi}+c^2(1+r_{s}^-)$. In this case the
simultaneous solution of Eqs. (\ref{00k}) and (\ref{kgk}) leads to
a potential that can be expressed as a series expansion in inverse
square potential. Approximating the potential by its leading term
we  write \be \n{pok} U(\phi)\approx
\frac{2\ga_\phi}{3\nu_\phi^2(1+r^-) F_x(-\phi_0^2)\phi^2}, \ee
where $
 \ga_\phi\approx 2\phi_0^2F_x(-\phi_0^2)/[F(-\phi_0^2)+ 2\phi_0^2F_x(-\phi_0^2)]$
along with the {\it k}-essence field, $\phi\approx  \phi_{0}\,t$.
When $\nu_\phi>0$ we apply a dual transformation to reverse its
sign.

\subsection{Tachyon field cosmology}
The energy density and
pressure of the phantom tachyon field
$\phi$ generated by the kinetic function $F(x)%
=(1+sx)^{1/2}=(1-s\dot\phi^2)^{1/2}$ are \cite{jlr} \be
\rho_{\phi}=U\left(1 -s\dot{\phi}^{2}\right)^{-1/2}\, , \quad
 p_{\phi} =- U \sqrt{1 -s \dot{\phi}^{2}} \, ,
\label{Trho&p} \ee
respectively, and its barotropic index is given by $\gamma_{\phi} = s%
\,\dot{\phi}^{2}$. A negative barotropic index is obtained in two
separate cases, viz, when $s<0$ with a real tachyon field and when
$s>0$ with an imaginary tachyon field.

Assuming an interaction between the tachyon field  and matter
governed by equations (\ref{consv}), with the subscript $\vphi$
replaced by $\phi$, and proceeding along parallel
lines to those sketched above one finds that the ratio $r =%
\rho_{m}/\rho_{\phi}$ evolves from $r^{+}_{s}$ to $r^{-}_{s}$ and
$a\propto t^{2/3\nu_\phi}$ where
$\nu_\phi=\gamma_{\phi}+c^2(1+r_{s}^-)$. In this case duality,
which requires that $\rho_\phi\to\rho_\phi$, $\ga_\phi\to
-\ga_\phi$ and $\Pi\to-\Pi$, leads to the following
transformations for the tachyon field and its potential
\\
\ben
\n{tfk} \dot{\phi}^2 \to
-\frac{s}{\bar s}\dot\phi^2, \qquad  U_0 \to
-\frac{s\sqrt{1+s\dot\phi_0^2}}{\bar
s\sqrt{1-s\dot\phi_0^2}}\,\,U_{0} \, .
\een
\\
As above, when $\nu_\phi>0$ we can apply a dual transformation to
reverse its sign.

\section{Concluding remarks}
We have considered a homogeneous, isotropic and spatially flat
universe dominated by two fluids (pressureless matter and dark
energy) that do not conserve separately but interact with each
other. We have shown that in this scenario there is a dual
symmetry transformation, given by $\rho_{i} \rightarrow \rho_{i}$,
$H \rightarrow -H$, $\gamma_{i} \rightarrow -\gamma_{i}$, and $\Pi%
\rightarrow - \Pi$, that preserves the form of Einstein's
equations irrespective of whether the dark energy is phantom  or
not. As a consequence, superaccelerated expansions can be obtained
from decelerated ones an viceversa without affecting the field
equations also in the case that matter and dark energy interact.

We observe, by passing, that if the interaction term is given by $\Pi =%
-c^{2}(\rho_{1}+ \rho_{2})$, then the cosmic coincidence problem
(i.e., ``why are the vacuum and dust energy densities of precisely
the same order today?" \cite{Paul}) is somewhat alleviated in the
sense that there is an attractor such that the energy densities of
matter and dark energy tend asymptotically to a fixed ratio,
$r_{s}^{-}$, regardless the dark energy component is phantom  or
not. Obviously, this does not solve the coincidence problem in
full. Its full solution would require to show, in addition, that
the attractor was reached only recently -or that we are very close
to it. Otherwise our approach would conflict with the tight
constraints imposed by the cosmic background radiation and the
standard scenario of large scale structure formation. On the other
hand, the precise value of $r_{s}^{-}$ cannot  be derived at
present. For the time being, it must be understood as an input
parameter. This is also the case of a handful of cosmic quantities
such as the present value  of the cosmic background radiation
temperature, or the ratio between the number of baryons and
photons.

\acknowledgments{This research has been partially supported by the
University of Buenos Aires and the ``Consejo Nacional de
Investigaciones Cient\'{i}ficas y T\'{e}cnicas"  under Projects
X224 and 02205, the Spanish Ministry of Science and Technology
under Grant BFM2003-06033, and the ``Direcci\'{o} General de
Recerca de Catalunya" under Grant 2005 SGR 00087.}

\end{document}